\documentclass[12pt]{article}
\usepackage{rotate}
\usepackage{epsfig}
\usepackage{psfrag}
\usepackage{a4}

\begin{document}

\newcommand \be  {\begin{equation}}
\newcommand \bea {\begin{eqnarray} \nonumber }
\newcommand \ee  {\end{equation}}
\newcommand \eea {\end{eqnarray}}

\title{\bf Random walks, liquidity molasses and critical response 
in financial markets}

\author{Jean-Philippe Bouchaud$^{*}$, Julien Kockelkoren$^{*}$, Marc Potters$^*$}
\maketitle
{\small
{$^*$ Science \& Finance, Capital Fund Management, 6-8 Bvd Haussmann}\\
{75009 Paris, France}
}

\begin{abstract}
Stock prices are observed to be random walks in time despite a strong, 
long term memory in the signs of trades (buys or sells). Lillo and
Farmer have recently suggested that these correlations are compensated by 
opposite long ranged fluctuations in liquidity, 
with an otherwise {\it permanent} market impact,
challenging the scenario proposed in 
Quantitative Finance {\bf 4}, 176 (2004), where the impact 
is {\it transient}, with a power-law decay in time. The exponent 
of this decay is precisely 
tuned to a critical value, ensuring simultaneously that prices are 
diffusive on long time scales and that the response function is nearly 
constant.
We provide new analysis of empirical data that confirm and make more 
precise our previous claims. We
show that the power-law decay of the bare impact function comes 
both from an excess flow of limit order opposite to the 
market order flow, and to a systematic anti-correlation of the bid-ask motion 
between trades, two effects that create a `liquidity molasses' which dampens 
market volatility.
\end{abstract}


\pagebreak

\section{Introduction}

The volatility of financial assets is well 
known to be too much large compared to the prediction of Efficient Market Theory \cite{Shiller} 
and to exhibit intriguing statistical anomalies, such as intermittency and long range memory (for recent reviews, see \cite{LeBaron,Rama,Book,BMD}). 
The availability of all trades and quotes on electronic markets makes it possible to analyze in
details the intimate mechanisms leading to these anomalies.  
In a previous paper \cite{QF}, we have proposed, based on empirical data, that the random walk 
nature of prices (i.e. the absence of return autocorrelations)
is in fact highly non trivial and results from a fine-tuned competition between liquidity 
providers and liquidity takers. In order not to reveal their strategy, liquidity takers must decompose 
their orders 
in small trades that are diluted in time over a several hours to several days. This creates long range 
persistence in the `sign' of the market orders (i.e. buy, $\varepsilon=+1$ or sell $\varepsilon=-1$) 
\cite{Hasbrouck,Hopman,QF,longmemory}. 
This persistence
should naively lead to a positive correlations of the returns and a super-diffusive behaviour of the 
price \cite{QF,longmemory}. 
However, liquidity providers act such as to create long range anti-persistence  in price changes: 
liquidity providers make their profit on the bid-ask spread but lose money when the price makes large 
excursions, in which case they sell low and have to buy high (or vice versa) for inventory reasons. 
Both 
effects rather precisely compensate and lead to an overall diffusive behaviour (at least to a first 
approximation),
such that (statistical) arbitrage opportunities are absent, as expected. 
We have shown in \cite{QF} that this picture allows one to understand the temporal structure of the 
market impact 
function (which measures how a given trade affects on average future prices), which was found to first increase, 
reach a maximum and finally decrease at large time, reflecting the mean-reversion action of 
liquidity providers.

The above picture was recently challenged by Lillo and Farmer \cite{longmemory}. Although they also find long memory (i.e., 
non summable power-law correlations) in the sign of market orders, 
they claim that the compensating mechanism
that leads to uncorrelated returns is not the slow, mean-reverting influence of 
liquidity providers suggested in \cite{QF}. Rather, they
argue that long range {\it liquidity fluctuations}, correlated with 
the order flow, act to suppress the otherwise permanent impact of
market orders and make the price diffusive. 

The aim of this paper is to explain in more details 
the differences and similarities between these conflicting pictures, 
and to present new data that support our original 
assertions \cite{QF}. While our 
previous paper mainly 
discussed on the case of France-Telecom, we also present a more systematic 
account of our main observables for
a substantial set of stocks from the Paris Bourse. 
We also give a much more precise qualitative
and quantitative description
of the way liquidity providers manage, on average, 
to mean-revert the price by monitoring 
the flow of limit orders. We therefore argue that liquidity providers create 
a kind of `liquidity molasses' that stabilises the volatility of financial markets,
which is indeed the traditional role given to market makers.     

\section{The impact of trades on prices}

\subsection{Formulation of the problem} 

In the following, we will consider follow the dynamics of prices in trade time $n$ (i.e. each distinct trade
increases $n$ by one unit) and define prices $p_n$ as the midpoint just before the $n^{th}$ trade: $p_n=(a_n+b_n)/2$,
where $a_n$ and $b_n$ are, respectively, the ask price and the bid price corresponding to the last quote before the trade. 
We assume that the price can be written in general as: 
\be\label{model}
p_n = \sum_{n'<n} {\cal G}\left(n,n'|\varepsilon_{n'},V_{n'},{\cal S}_{n'}\right) 
\ee
where ${\cal G}$ describes the impact at time $n$ of a trade at time $n'$, of sign and volume $\varepsilon_{n'},V_{n'}$,
knowing that the order book at time $n'$ is in a certain state ${\cal S}_{n'}$ (specified by the list of all prices and
volumes of the limit orders). The assumption we made in \cite{QF} is that the impact function ${\cal G}$ can be 
decomposed into an average, systematic part in the direction of the trade, plus fluctuations: 
\be
{\cal G}\left(n,n'|\varepsilon_{n'},V_{n'},{\cal S}_{n'}\right)\equiv \varepsilon_{n'} G(n,n'|V_{n'}) + 
\xi(n,n'),
\ee
where the function $G$ was furthermore assumed to by time translation invariant\footnote{This is probably only an approximation 
since time of the day, for example, should matter.} 
and factorisable as: $G(n,n'|V_{n'})= f(V_n) G_0(n-n')$. The last assumption 
is motivated by theoretical and empirical results \cite{Farmer1,Bali,QF}, where $f(V)$ is found to be a power-law
with a small exponent $f(V) \sim V^\beta$ \cite{Lillo,Lillo2} or a logarithm $f(V) \sim \ln V$ \cite{Bali,QF}. 
The noise term $\xi(n,n')$ is uncorrelated with the $\varepsilon_{n'}$ and has a variance $(n-n')D$. 
The final form of the model proposed in \cite{QF} therefore reads: 
\be\label{modelQF}
p_n = \sum_{n'<n} G_0(n-n') \varepsilon_{n'} \ln V_{n'} + \xi(n,n').
\ee
The main finding of \cite{QF} is that the bare impact function $G_0(\ell)$ must decay with the time lag in order
to compensate for the long range correlation in the $\varepsilon$, in other words that the impact of a single trade 
is transient rather than permanent. In their recent work, Lillo and Farmer \cite{longmemory} argue that it is
rather the 
fluctuations in liquidity (encoded in the instantaneous shape of the order book ${\cal S}_{n'}$), that are crucial. 
Their model amounts to write $p_n$ as:
\be\label{modelLF}
p_n = \sum_{n'<n} \frac{\varepsilon_{n'} V_{n'}^\beta}{\lambda({\cal S}_{n'})} + \xi(n,n'),
\ee
with $\beta=0.3$ and where $\lambda$ is the instantaneous liquidity of the market. The difference 
between $V^{.3}$ and $\ln V$ is actually not relevant; rather, the crucial difference between Eq. (\ref{modelQF}) and Eq. 
(\ref{modelLF}) is that the impact is {\it transient} in the former case and {\it permanent} (but fluctuating) in the 
latter case, a point on which we will comment later. 

The argument of ref. \cite{longmemory} in favor of the second model, Eq. (\ref{modelLF}) goes in two steps: first, they
propose, as a proxy of the instantaneous liquidity $\lambda_n$, the volume $v_n$ at 
the best price (i.e. ask for buys and bid for sells): see  \cite{longmemory} section VI B.
They then study the time series of $r_n={\varepsilon_{n} V_{n}^\beta}/{v_n}$ and find that linear correlations 
have nearly completely disappeared, at variance with the unrescaled series ${\varepsilon_{n} V_{n}^\beta}$ that exhibit the
problematic long range correlations. Their conclusion is therefore that ``the inclusion of the time varying liquidity term 
apparently removes long-memory''. Here, we want to refute this interpretation based on 
three independent sets of arguments:
a) we show that Eq. (\ref{modelLF}) has less explicative power than Eq. (\ref{modelQF}); 
b) Eq. (\ref{modelLF}) leads to an 
average response function (see \cite{QF} and below) 
that significantly increases with time lag, at variance with data
and c) the absence of linear correlations observed in $r_n$ is an artefact coming 
from the very large fluctuations 
of the volume at the best price. Note that our data concerns stocks from
Paris Bourse rather than the LSE stocks studied in \cite{longmemory}. However,
we do not expect major qualitative differences between the two markets. 

\subsection{Response functions}

We first start by recalling the definition of the average response function, as the 
correlation between the
sign of a trade at time $n$ and the subsequent price difference between $n$ and $n+\ell$ \cite{QF}:
\be
{\cal R}(\ell) = \left \langle \left(p_{n+\ell}-p_n\right) \cdot \varepsilon_n \right \rangle,
\ee
The quantity ${\cal R}(\ell)$ measures how much, on average, the price moves up 
conditioned to a buy order at time $0$ (or how
a sell order moves the price down) a time $\ell$ later. Note that because of the temporal correlations 
between the $\varepsilon$'s, this quantity is {\it not} the above market response to a single trade $G_0(\ell)$ \cite{QF}.
This quantity is plotted in Fig. 1 for Carrefour in 2001, 2002. 
As emphasized in \cite{Bali,QF}, ${\cal R}(\ell)$ 
is found to weakly increase up to a maximum beyond which it decays back and can even change sign for large $\ell$ 
(see Figs. 2, 3). For other
stocks, or other periods, the maximum of ${\cal R}(\ell)$ is not observed, and 
${\cal R}(\ell)$ is seen to 
increase (although always rather mildly, at most by a factor 3) with $\ell$: 
see Figs. 2,3. As will 
be clear below, this difference of behaviour can actually be understood 
within our model.

\begin{figure}
\begin{center}
\psfig{file=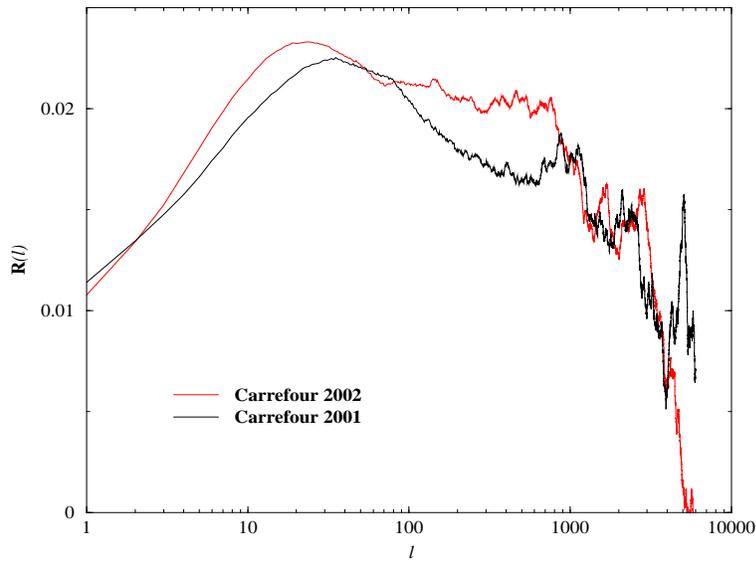,width=7.5cm,height=10cm,angle=270} 
\end{center}
\caption{Response function ${\cal R}(\ell)$ (in Euros) for Carrefour in the periods 2001 and 2002.}
\label{Fig1}
\end{figure}

\begin{figure}
\begin{center}
\psfig{file=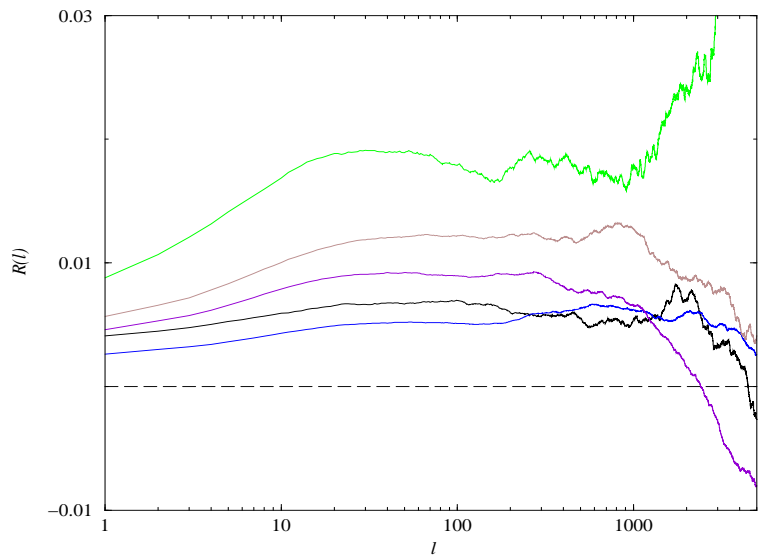,width=7.5cm,height=10cm,angle=270} 
\end{center}
\caption{Response function ${\cal R}(\ell)$ (in Euros) for stocks from Paris Bourse in 2002. From top to bottom: EN, EX, FTE, ACA, CGE.
(See Table 1 for the stocks code). Note that for some stocks ${\cal R}(\ell)$ increases for all $\ell$
(see e.g. CGE), whereas for other stocks ${\cal R}(\ell)$ reaches a maximum before becoming
negative (see e.g. ACA). The dotted line correspond to  ${\cal R}(\ell)=0$.
}
\label{Fig2a}
\end{figure}

\begin{figure}
\begin{center}
\psfig{file=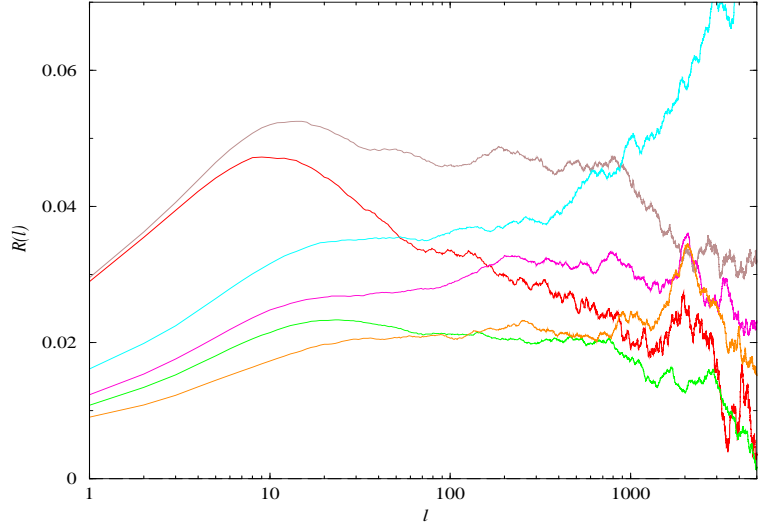,width=7.5cm,height=10cm,angle=270}
\end{center}
\caption{Response function ${\cal R}(\ell)$ (in Euros) for other stocks from Paris Bourse in 2002. From top to bottom: FP, BN, GLE, MC, CA, VIE.
(See Table 1 for the stocks code). Note that for some stocks ${\cal R}(\ell)$ increases for all $\ell$
(see e.g. GLE), whereas for other stocks ${\cal R}(\ell)$ reaches a maximum before becoming
negative (see e.g. CA, for $\ell > 5000$). 
}
\label{Fig2b}
\end{figure}

\begin{table}
\begin{center}
\begin{tabular}{||l|c|c|c|c|c||} \hline\hline
Code\ \hspace{0.1cm} \   & 
\hspace{0.1cm}  Stock name \hspace{0.1cm} &
\hspace{0.1cm} Av. price \hspace{0.1cm} &
\hspace{0.1cm} Av. tick  \hspace{0.1cm} &
\hspace{0.1cm} Av. spread \hspace{0.1cm} &
\hspace{0.1cm} \# trades \hspace{0.1cm} 
\\ \hline
ACA  &  Cr\'edit Agricole &  19.63  &  0.01 &  0.0408 & 379,000  \\ \hline
BN  &  Danone &   132.50 &  0.1 &  0.154 & 351,000 \\ \hline
CA  &  Carrefour & 48.54  &  0.0268 &  0.0578 & 555,000 \\ \hline
CGE  &  Alcatel & 9.85  & 0.01 &  0.015 & 1,020,000 \\ \hline
EN  &  Bouygues & 29.69 & 0.01 &  0.0413 & 240,000 \\ \hline
EX  &  Vivendi & 27.47  &  0.0126 &  0.0287 & 979,000 \\ \hline
FP  &  Total & 152.27  & 0.1 &  0.136 & 759,000 \\ \hline
FTE  &  France-Telecom &  21.04 &  0.01 & 0.024 & 1,051,000 \\ \hline
GLE  &  Soci\'et\'e G\'en\'erale & 61.80  & 0.043 &  0.0735 & 499,000 \\ \hline
MC  &  LVMH & 47.71 & 0.0209 & 0.0566 & 437,000 \\ \hline
VIE  &  Vivendi Env. & 29.75 & 0.01 &  0.0452 & 226,000 \\ \hline\hline
\end{tabular}
\end{center}
\caption[]{\small Selection of stocks studied in this paper, with the
average price, tick size and bid-ask spread in Euros in 2002. We also give
the total number of trades in 2002. The results reported here 
qualitatively hold for most other stocks from Paris Bourse, 
but also other exchanges (see
\cite{QF,longmemory}).
}
\end{table}

In Fig. 4, we also plot three other, similar quantities. 
The first is the (normalized) 
correlation between the price change and $\varepsilon_n \ln V_{n}$:
\be
{\cal R}_V(\ell) = \frac{\left \langle \left(p_{n+\ell}-p_n\right) \cdot 
\varepsilon_n \ln V_n \right \rangle}
{\langle \ln^2 V_n \rangle^{1/2}}
\ee
which has a similar shape but is distincly larger than ${\cal R}$ itself, 
showing that, as expected, the variable
$\varepsilon_n \ln V_n$ has a larger explicative power than $\varepsilon_n$ itself. 
In order to test the Lillo-Farmer 
model, we have also computed two further quantities. One is the normalized 
correlation between the Lillo-Farmer variable $r_n={\varepsilon_{n} V_{n}^\beta}/{v_n}$ 
and the empirical price change:
\be
{\cal R}_{LF}(\ell) = 
\frac{\left \langle \left(p_{n+\ell}-p_n\right) \cdot r_n  \right \rangle}
{\langle r_n^2 \rangle^{1/2}}.
\ee
This quantity  measures the explicative power of $r_n$, and
can be directly compared to ${\cal R}$ and ${\cal R}_V$.  
As can be seen in Fig. 4, ${\cal R}_{LF}(\ell)$ is in fact 
a factor 3 smaller than ${\cal R}_V(\ell)$ (see also the quantity $Z$ in Table 2, 
last column). 

The second interesting quantity is: 
\be
{\cal R}^*_{LF}(\ell) = \left \langle \left(\sum_{n'=n}^{n+\ell-1} r_{n'} \right) \cdot \varepsilon_n \right \rangle.
\ee

\begin{figure}
\begin{center}
\psfig{file=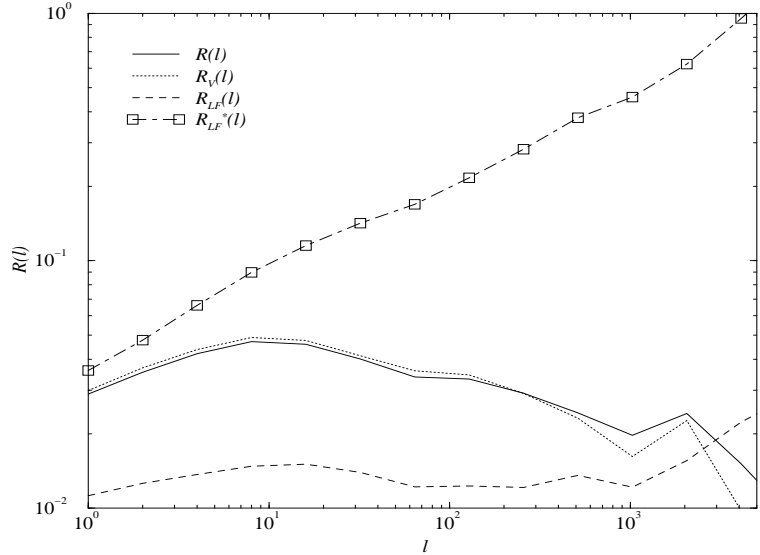,width=7.5cm,height=10cm,angle=270} 
\end{center}
\caption{Four different `response functions' ${\cal R}(\ell)$, ${\cal R}_V(\ell)$, ${\cal R}_{LF}(\ell)$ 
and ${\cal R}^*_{LF}(\ell)$, (see text) in Euros for BN in 2002. This plot shows (a) that the Lillo-Farmer
variable $r_n$ has a weak explicative power (see ${\cal R}_{LF}$ -- dashed line) 
and (b) that their permanent impact model leads to a considerable
over-estimation of the true response function (see ${\cal R}^*_{LF}$ -- dashed-dotted lines, showing a 30 times increase with $\ell$).}
\label{Fig3}
\end{figure}

The quantity measures a fictitious average response function, which would 
follow if the price dynamics was given by Eq. (\ref{modelLF}).
We see in Fig. 4 that ${\cal R}^*_{LF}(\ell)$, 
at variance with the true ${\cal R}(\ell)$, sharply grows with $\ell$, 
as a consequence of the correlation of the $\varepsilon$'s which 
are not compensated by a fluctuating liquidity.
As we have mentioned in \cite{QF}, the response function ${\cal R}(\ell)$ 
is a very sensitive measure of the
dynamics of prices that allows one to reveal subtle effects, beyond 
the simple autocorrelation of price changes (see also below).

Finally, we show in Fig. 5 the rapid fall of 
the autocorrelation of the variables $r_n$, that was argued by Lillo and 
Farmer to be a strong support to their model \cite{longmemory}.
Unfortunately, this effect is not relevant and 
is due to the fact that 
the volume at the best price has large fluctuations. For example, 
in the case of FTE, the distribution of $v$ is found to be well-fit by
$P(v) \propto v^{\mu-1} \exp(-v/v_0)$ with $\mu > 1$, so that 
the most probable values correspond to $v \sim 1$, whereas the mean value 
is $\sim 3000$ \cite{BMP}. 
Since $v_n$ appears in the denominator of $r_n$, it is
clear that the $r_n$ correlations are dominated by times where the 
volume at bid/ask is particularly small; these small values show 
little autocorrelations (see Fig. 5).\footnote{After discussions, Lillo and
Farmer have agreed that their results on LSE stocks are in fact compatible
with the above interpretation.}

\begin{figure}
\begin{center}
\psfig{file=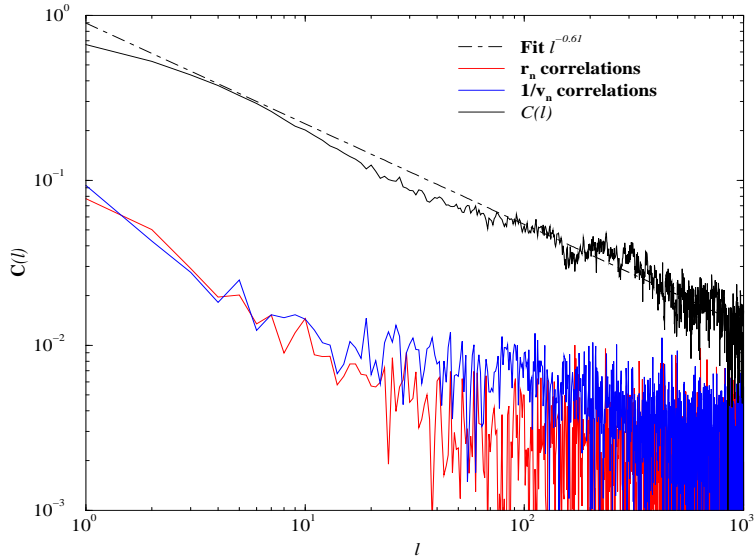,width=7.5cm,height=10cm,angle=270} 
\end{center}
\caption{Sign correlations ${\cal C}(\ell)$ for BN, showing a long range, 
power-law decay, and comparison between the smaller and faster decaying correlation of the $r_n$ and
the $1/v_n$, showing that the former is dominated by the weak correlations 
between small order volumes, and not by a compensation between market order
flows and limit order flows.}
\label{Fig4}
\end{figure}

\subsection{The bare impact function and price diffusion}

We conclude from Fig. 4 that although the variables $r_n$ are indeed close to 
being uncorrelated, they do not provide an 
adequate basis to interpret the dynamics of real price. 
Our transient impact model, on the other 
hand, allows one
to reconcile the absence of autocorrelations in price changes with the observed non monotonous shape of 
the average
response function, provided the bare impact function $G_0(\ell)$ is chosen adequately. In \cite{QF}, 
it was shown that
if the correlation of the $\varepsilon$'s decays as $\ell^{-\gamma}$, then $G_0(\ell)$ should also 
decay, at large times,
as a power-law $\ell^{-\beta}$ with $\beta \approx (1-\gamma)/2$. For $\beta > (1-\gamma)/2$, the 
price is subdiffusive
(anti-persistent) and the response function ${\cal R}(\ell)$ has a maximum before becoming negative 
at large $\ell$. 
For $\beta < (1-\gamma)/2$, on the other hand, the price is superdiffusive (persistent) and the 
response function 
monotonously increases (see Fig. 10 of \cite{QF}). The short time behaviour of $G_0(\ell)$ can in 
fact be extracted 
from empirical data by using the following exact relationship:
\be \label{response}
{\cal R}(\ell) = \langle \ln V \rangle G_0(\ell)+ 
\sum_{0 < n < \ell} G_0(\ell-n) {\cal C}(n) + 
\sum_{n > 0} \left[G_0(\ell+n)-G_0(n)\right] {\cal C}(n).
\ee
where:
\be
{\cal C}(\ell)= \langle \varepsilon_{n+\ell} \,\, \varepsilon_n \ln V_n \rangle,
\ee
a correlation function that can also be measured directly (see Figs. 5,6). 

\begin{figure}
\begin{center}
\psfig{file=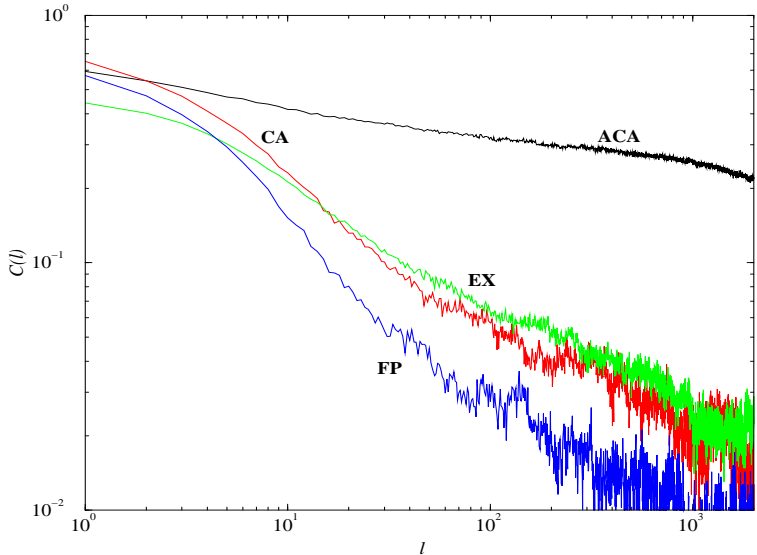,width=7.5cm,height=10cm,angle=270} 
\end{center}
\caption{Plot of the sign correlations ${\cal C}(\ell)$ for a selection of four stocks, 
showing the long-ranged nature of these correlations. See also Table 3.}
\label{Fig5}
\end{figure}

Eq. (\ref{response}) gives a set of linear 
equations relating ${\cal R}$, $G_0$ and ${\cal C}$ that can easily be solved for $G_0$. The result is 
plotted in Fig. 7 for different stocks. One sees that $G_0(\ell)$ 
is first flat or rises very slightly with 
$\ell$ before indeed 
decaying, for $\ell \gg 1$, like a power law, with 
$\beta$ given in Table 2. The fit used to extract the value of $\beta$ is
$G_0^f(\ell)=\Gamma_0/(\ell_0^2+\ell^2)^{\beta/2}$ which is similar, but
not identical to, the one proposed in \cite{QF}. The advantage of the present 
fit is that it matches quite well the rather flat initial behaviour of 
$G_0(\ell)$. We also give in Table 2 the value of 
other quantities such as the exponent $\gamma$ governing 
the decay of the $\varepsilon$ correlations.
A very similar shape for $G_0$ can be observed for all stocks; 
fluctuations around the critical line
$\beta = (1-\gamma)/2$ (see Fig. 8) are enough to explain the fact that 
${\cal R}$ sometimes has a 
maximum, sometimes not. 

\begin{figure}
\begin{center}
\psfig{file=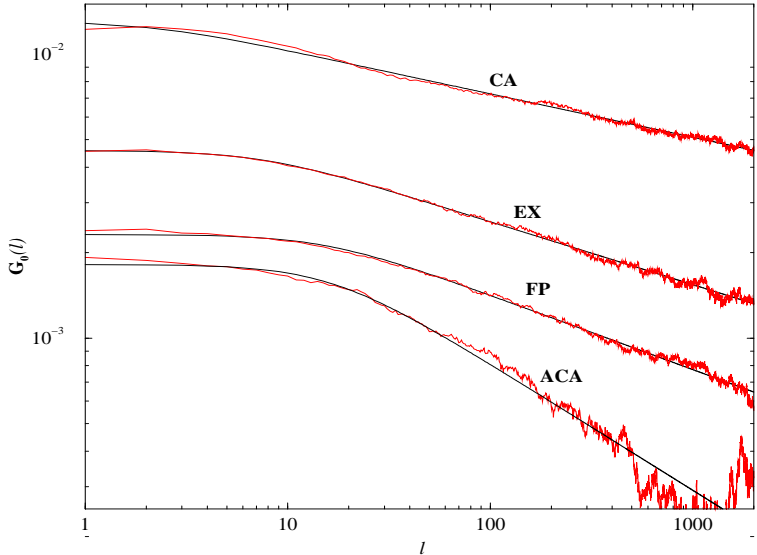,width=7.5cm,height=10cm,angle=270} 
\end{center}
\caption{Comparison betwen the empirically determined $G_0(\ell)$, extracted from ${\cal R}$ and
${\cal C}$ using Eq.(\ref{response}),
and the fit 
$G_0^f(\ell)=\Gamma_0/(\ell_0^2+\ell^2)^{\beta/2}$, used to extract the parameters given in
Table 2, for a selection of four stocks: ACA, CA, EX, FP.}
\label{Fig7}
\end{figure}

\begin{figure}
\begin{center}
\psfig{file=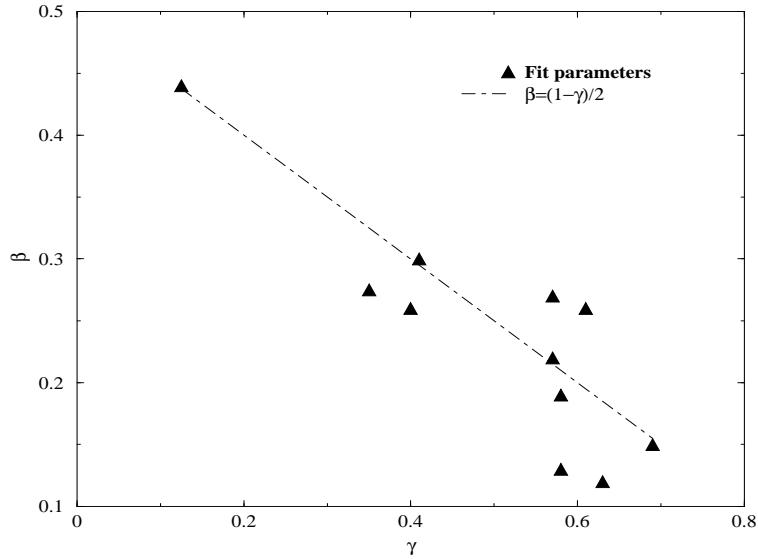,width=7.5cm,height=10cm,angle=270} 
\end{center}
\caption{Scatter plot of the exponents $\beta, \gamma$ extracted from the fit of $G_0$ and
$\cal C$. These exponents are seen to lie in the vicinity of the critical line $\beta=(1-\gamma)/2$ 
(dotted line), as expected from the nearly diffusive behaviour of prices (see Fig. 9), and \cite{QF}.}
\label{Fig8}
\end{figure}

Correspondingly, the vicinity of the critical line ensures that the price has a diffusive 
behaviour, as is indeed confirmed by measuring the variance of price changes:
\be
{\cal D}(\ell) = \langle \left(p_{n+\ell}-p_n\right)^2 \rangle 
\approx D \ell; \qquad \forall \ell,
\ee
as demonstrated in Figs. 9 and 10. The fact that ${\cal D}(\ell)$ is strictly linear in
$\ell$ is of course tantamount to saying that price increments are uncorrelated.

\begin{table}
\begin{center}
\begin{tabular}{||l|c|c|c|c|c|c|c||} \hline\hline
Stock\ \hspace{0.2cm} \   & 
\hspace{0.2cm} $\sqrt{{\cal D}(1)}$ \hspace{0.2cm} &
\hspace{0.2cm}  $\Gamma_0$ \hspace{0.2cm} &
\hspace{0.2cm}  $\ell_0$ \hspace{0.2cm} &
\hspace{0.2cm}  $\beta$ \hspace{0.2cm} &
\hspace{0.2cm}  $C_0$ \hspace{0.2cm} &
\hspace{0.2cm}  $\gamma$ \hspace{0.2cm}&
\hspace{0.2cm}  $Z$ \hspace{0.2cm}
\\ \hline
ACA  &  1.69 &  0.63 & 16.3 & 0.44 & 0.58 & 0.125 & 0.35 \\ \hline
BN  &  7.9 &  1.75 &  3.1 & 0.26 & 0.81 & 0.61 & 0.37 \\ \hline
CA  &  3.13 &  0.71 & 7.4 & 0.22 & 0.83 & 0.57 & 0.27 \\ \hline
CGE  &  0.84 &  0.20 & 8.9 & 0.275 & 0.49 & 0.35 & 0.18 \\ \hline
EN  &  2.75 &  0.66 & 9.2 & 0.27 & 0.83 & 0.57 & 0.27 \\ \hline
EX  &  1.79 &  0.47 & 15.3 & 0.26 & 0.45 & 0.40 & 0.20\\ \hline
FP  &  7.0 &  1.46 & 2.2 & 0.15 & 0.79 & 0.69 & 0.28 \\ \hline
FTE  &  3.9 &  0.47 & 20.3 & 0.30 & 0.52 & 0.41 & 0.23 \\ \hline
GLE  &  4.37 &  0.73 & 0.7* & 0.13 & 0.86 & 0.58 & 0.28 \\ \hline
MC  &  3.47 &  0.67 & 3.1 & 0.19 & 0.95 & 0.58 & 0.26 \\ \hline
VIE  &  2.8 &  0.38 & 0.25* & 0.12 & 0.75 & 0.63 & 0.26 \\ \hline\hline
\end{tabular}
\end{center}
\caption[]{\small Summary of the different quantities and fit parameters 
for 11 stocks of the Paris Bourse during the year 2002. 
$G_0(\ell)$ is fitted as: 
$G_0(\ell)=\Gamma_0/(\ell_0^2+\ell^2)^{\beta/2}$, and 
${\cal C}(\ell)=C_0/\ell^\gamma$, both in the 
range $\ell=2 \to 2000$. $\sqrt{D(1)}$ and $\Gamma_0$ are in cents of Euros.
The * means that the fit of $G_0$ for small $\ell$ is not very good.
The last column measures the {\it relative} explicative power of the Lillo-Farmer
variable, compared to our own: $Z={\cal R}_{LF}(1)/{\cal R}(1)$. 
}
\end{table}

\begin{figure}
\begin{center}
\psfig{file=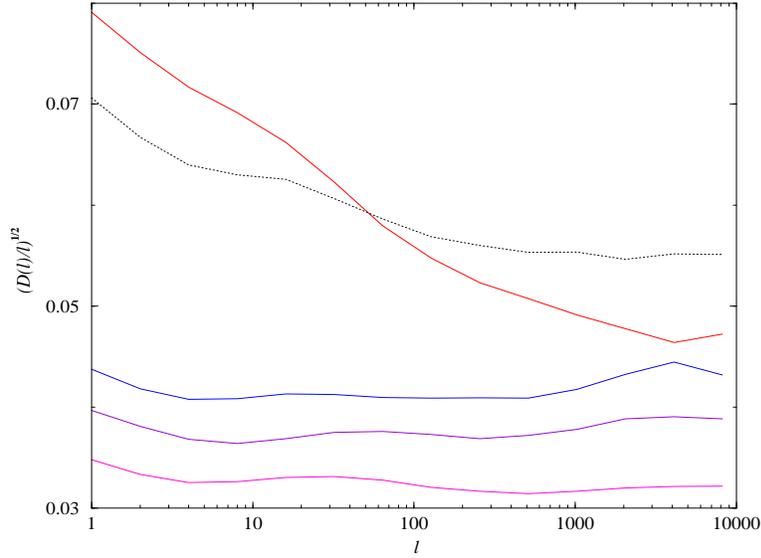,width=7.5cm,height=10cm,angle=270} 
\end{center}
\caption{Plot of $\sqrt{{\cal D}(\ell)/\ell}$ (in Euros) vs. 
$\ell$ for several stocks. Apart 
from BN and FP (for which the tick size is large), this quantity 
is roughly constant with $\ell$, 
showing that prices are to a very good 
approximation diffusive, even on shortest times scales. From top to bottom: 
BN, FP, GLE, FTE, MC.}
\label{Fig9a}
\end{figure}

\begin{figure}
\begin{center}
\psfig{file=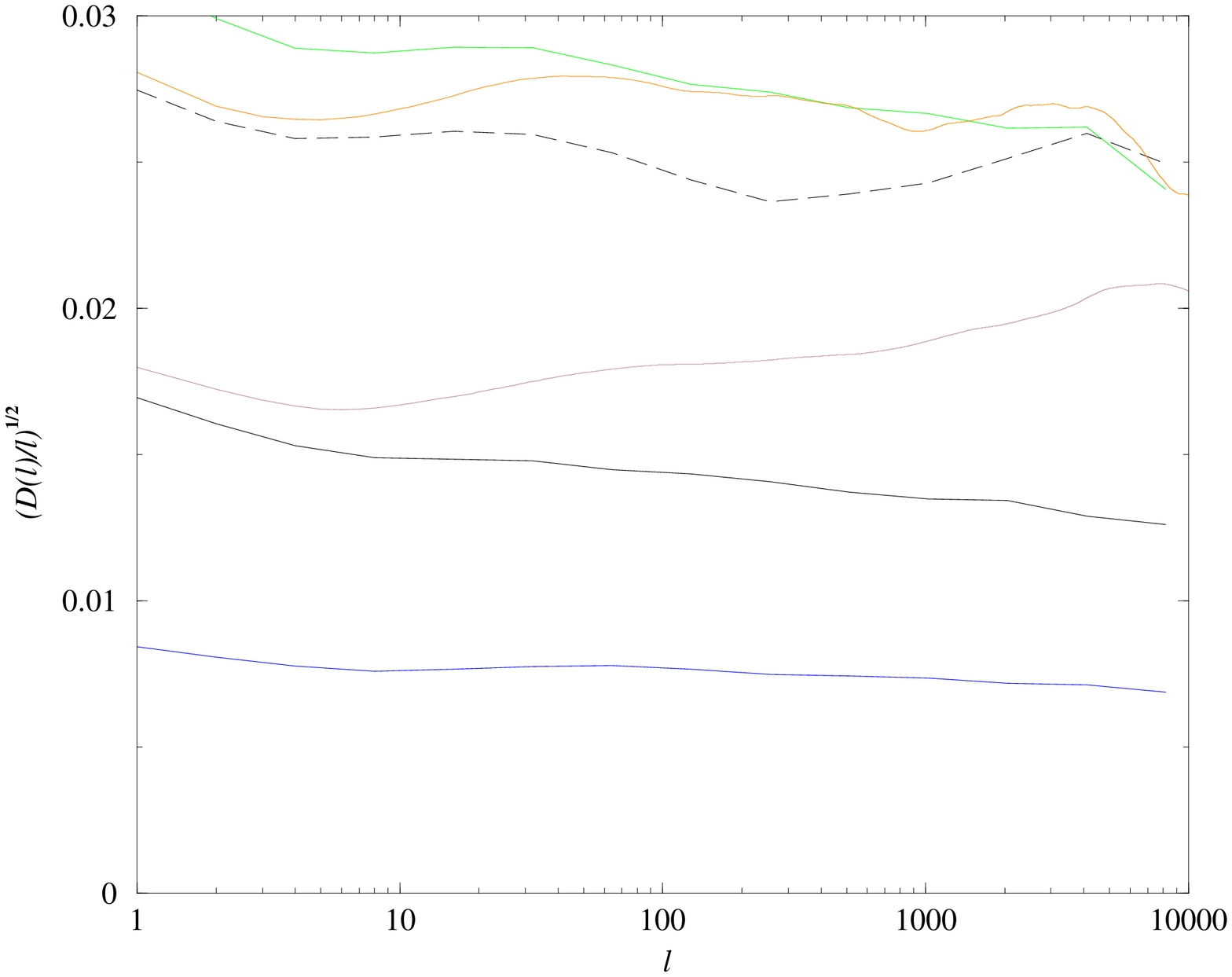,width=7.5cm,height=10cm,angle=270}
\end{center}
\caption{Plot of $\sqrt{{\cal D}(\ell)/\ell}$ (in Euros) vs. $\ell$ for 
all other (smaller tick) stocks. From top to bottom: CA, VIE,
EN, EX, ACA, CGE.}
\label{Fig9b}
\end{figure}

\subsection{Economic interpretation of the shape of the bare impact} 

The economic interpretation of the non monotonic behaviour of $G_0$ is as follows. Suppose that you are a liquidity 
provider, making profits on the bid-ask spread and losses on large price excursions, and that you see a flow of
buy orders coming. In the absence of news and for typical buy volumes,\footnote{The following discussion 
is intended to describe typical situations. Obviously, if the buy volume is anomalously large, liquidity
providers would anticipate some insider information and react differently.} the natural strategy is, on short times, 
to biais the ask price up to be able to sell higher 
while there are clients eager to buy. However, you now have a net short position on the stock that you want to eventually 
shift back to zero. So you would like to buy back, in the near future, 
at the cheapest possible price. 
In order to prevent the price from going up, 
you can/should do two things: a) create a barrier to further price 
rises by placing a large number of sell orders at 
the ask, off which the price will bounce back down b) 
place bid orders as low as possible. 
Both effects act to create a liquidity molasses that mean revert the price 
towards its initial value. {\it Both} these effects can actually be 
observed directly on the data.

\begin{itemize}

\item a) One observes a strong correlation between a buy (resp. sell) market order 
moving the price up and the subsequent appearance of limit orders at the ask (resp. bid) 
\cite{Rosenownew,longmemory}. If a `wall' of limit orders appears at the ask while the  
bid remains poorly populated, the probability that the price moves down upon the arrival 
of further market orders becomes larger than the probability to move up. One can
visualize this effect more clearly by separating the total
price change into two components: price variations due to market orders, $\Delta_M p_n$, 
corresponding to the change of mid-point between the quote immediately prior and the 
quote immediately posterior to the $n$-th trade, and price variations due to limit orders,
 $\Delta_L p_n$ corresponding to changes of mid-points in-between trade $n$ and trade
$n+1$. By definition, 
\be
p_{n+\ell}-p_n = \sum_{k=n}^{n+\ell-1} \left[\Delta_M p_k + 
\Delta_L p_k\right] \equiv \left(p_{n+\ell}-p_n\right)_M + 
\left(p_{n+\ell}-p_n\right)_L.
\ee 
One can then measure the response function 
restricted to price changes due to market orders:
\be
{\cal R}_M(\ell) = \left \langle \left(p_{n+\ell}-p_n\right)_M \cdot 
\varepsilon_n \right \rangle,
\ee
and compare it (see Fig. 11) to ${\cal R}(\ell)$. We observe for all
stocks that ${\cal R}_M(\ell)$ 
and ${\cal R}(\ell)$ have the same overall shape. For FTE, for example, 
${\cal R}_M(\ell)$ 
also bends down and becomes negative for large $\ell$. But since by definition
$\Delta_M p_k = \varepsilon_k {\cal G}_k$ with ${\cal G}_k \ge 0$ (a buy market order
can only move the price up or leave it unchanged), the fact that ${\cal R}_M(\ell)$ 
decreases implies that ${\cal G}_k$ is {\it anticorrelated} with $\varepsilon_n 
\varepsilon_k$. In other words, sell orders following buy orders impact the price 
more than buy orders following buy orders, as expected if the order book fills 
in more on the ask side than on the bid side after a buy market order (and, of
course, similarly for the sell side).     

\item b) there is an anticorrelation between buy orders and the subsequent motion of 
the bid-ask {\it in-between} trades. This is seen both from the fact that 
${\cal R}_M(\ell) > {\cal R}(\ell)$ for $\ell$ not too large (see Fig. 11), implying that the 
response function restriced to limit orders is negative. Furthermore, one can 
study the correlation between a market order induced price change $\Delta_M p_n$ 
and a later limit order price change $\Delta_L p_{n+\ell}$, which is found to
be negative (as also reported in \cite{QF,longmemory}). This compensates the 
positive correlations between $\Delta_M p_n$ and  $\Delta_M p_{n+\ell}$ (and
between $\Delta_L p_n$ and $\Delta_L p_{n+\ell}$), that would otherwise lead 
to a superdiffusion in the price. 

\end{itemize}

\begin{figure}
\begin{center}
\psfig{file=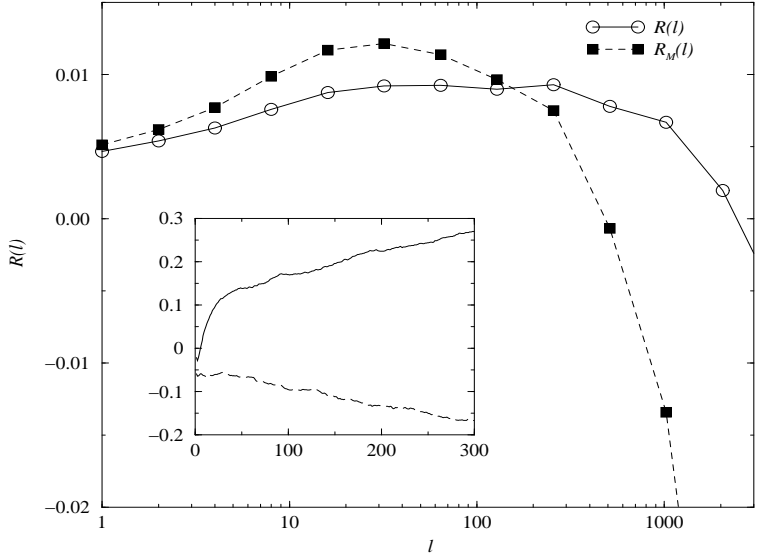,width=7.5cm,height=10cm,angle=270} 
\end{center}
\caption{Main figure: Comparison between the full response $\cal R$ (circles) 
and the response restricted to market order induced price changes ${\cal R}_M$ (squares),
for FTE in 2002. Inset: Integrated correlation functions, corresponding to
$\langle \Delta_M p_n \cdot \Delta_M p_{n+\ell}\rangle$ (full line), and 
$\langle \Delta_M p_n \cdot \Delta_L p_{n+\ell}\rangle$ (dotted line). The former
is clearly positive, and is compensated by the negative correlation between
market orders induced shifts and subsequent changes in the mid-quotes.}
\label{Fig10}
\end{figure}

In order to make our point even more clearly, it is useful to emphasize 
the antagonist forces present in financial markets: 
\begin{itemize}

\item The ideal world 
for liquidity providers is a stable, fixed average price
that allows them to earn the bid-ask spread at every round-turn. 
Volatility is the enemy\footnote{Insider information is also the 
liquidity provider enemy, 
but this situation is rather rare on the scale of the thousands of trades happening 
every day on each single liquid stock. However, creating a liquidity wall is indeed 
risky for the liquidity provider in the case where some true information motivates the 
market orders. In that case, the insider can use his information without 
impacting the price.}, liquidity molasses is the solution: a vanishing long term impact 
(i.e. $G_0(\infty)= 0$) is a way to limit the volatility of the market and 
to increase the liquidity provider gains. Reducing the volatility of financial
markets is in fact the traditional role given to market makers in 
non electronic markets. Note that we do not assume any kind of collusion 
between liquidity providers: they all, individually, follow a perfectly reasonable
strategy. 

\item Conversely, {\it permanent} impact is what the 
liquidity taker should hope for: if the price rises because of 
his very trade but stays high until he sells back, his impact is not really a cost.
On the other hand, if the price deflates back after having bought it, it means that 
he paid to much for it.\footnote{{\it The salesman knows 
nothing about what he is selling, save that he is charging a great deal too 
much for it.} (Oscar Wilde)} The correlations created by splitting his bid in small
quantities also help keeping the price up. 
\end{itemize}

These are the basic ingredients ruling 
the competition between liquidity providers and liquidity takers. The subtle balance 
between the positive correlation in the trades (measured by $\gamma$) and the 
liquidity molasses induced by liquidity providers (measured by $\beta$) 
is a {\it self-organized dynamical equilibrium}. Its stability 
comes from two counter-balancing effects: if the liquidity providers 
are too slow 
to revert the price ($\beta < (1-\gamma)/2$),
then the price is superdiffusive and liquidity providers lose money on 
average \cite{WB}; therefore they increase $\beta$. 
If the mean reversion is too strong ($\beta > (1-\gamma)/2$),  
the resulting long term anticorrelations is an incentive for buyers 
to wait for prices to come back down to continue buying. Liquidity takers 
thereby spread their trading over longer time scales, which corresponds to
smaller values of $\gamma$. 

A dynamical equilibrium where 
$\beta \approx (1-\gamma)/2$ therefore establishes 
itself spontaneously, with clear economic forces driving the 
system back towards this equilibrium. Interestingly, {\it fluctuations} 
around this critical line leads to fluctuations 
of the local volatility, since persistent patches correspond to high local 
volatility and antipersistent patches to low local volatility 
(see also \cite{Wyart} for a similar mechanism). Extreme crash situations 
are well-known to be liquidity crisis, where the liquidity molasses effect disappears 
temporarily, destabilising the market (on that point, see the detailed recent 
study of \cite{Farmernew,Rosenownew2}). 

Finally, the mean-reverting nature of the response function is of crucial importance
to understand the influence of volume and execution time on the actual impact 
of trading on prices (on this point, see \cite{Gabaix,GabaixFarmer}).

\section{Summary and Conclusion}

The aim of this paper was to challenge Lillo and Farmer's suggestion that 
the strong memory in the signs of trades is compensated by liquidity fluctuations, 
with an otherwise {\it permanent} market impact, and confirm
the more subtle scenario proposed in our previous paper \cite{QF}, in which the impact 
is {\it transient}, with a power-law decay in time. The exponent is precisely 
tuned to a critical value, ensuring simultaneously that prices are 
diffusive on long time scales and that the response function is nearly 
constant. 
Therefore, the seemingly 
trivial random walk behaviour of price changes in fact results from a 
fined-tuned competition between two opposite effects, 
one leading to super-diffusion 
-- the autocorrelation of market order flow; 
the other leading to sub-diffusion 
-- the decay of the bare impact function, 
reflecting the mean-reverting nature of the limit order flow. 
We have shown that
mean reversion comes both from an excess flow of limit order opposite to the 
market order flow, and to a systematic anti-correlation of the bid-ask motion 
between trades. Note that in the above picture, the random walk 
nature of prices and their volatility are induced by the trading 
mechanisms alone, with no reference to real news. These of course 
should also play a role, but probably not as important as pure
speculation and trading that lead to excess volatility (see the discussion
and references in
\cite{QF}).

The above fine tuning is however, obviously, not always perfect, 
and is expected to be only approximately true on average.
Breakdown of the balance between the two effects can lead either to large 
volatility periods and crashes when the liquidity
molasses disappears, or to 
low volatility periods when mean-reverting effects are strong. 
The small imbalance between the two effects therefore leads to 
different shapes of ${\cal R}(\ell)$ 
(monotone increasing or turning round and changing sign).
As emphasized in \cite{QF}, our finding that the absence of 
arbitrage opportunities results from a critical balance 
between antagonist effects might justify several claims made 
in the (econo-)physics literature that 
the anomalies in price statistics (fat tails in returns 
described by power laws \cite{Lux,Stanley}, 
long range self similar volatility correlations \cite{Rama,BMD}, and the long ranged 
correlations in signs \cite{QF,longmemory}) are due to 
the presence of a critical point in the vicinity of which the market operates 
(see e.g. \cite{soc}, and in the context of financial markets \cite{MG,lux,QFR}). From a more practical point of view, we hope that the present 
detailed picture of market microstructure could help understanding 
the mechanisms leading to excess volatility, and suggest ways to control more 
efficiently the stability of financial markets.   

\vskip 1cm
\section*{Acknowledgments} We want to 
thank Matthieu Wyart and Yuval Gefen for many inspiring discussions and
ideas about this work.  
We also thank Doyne Farmer and Fabrizio Lillo for many comments and e-mail exchanges
that allowed to clarify a lot the present paper.


\end{document}